\documentclass[useAMS,usegraphicx,usenatbib]{mn2e} % for the long lists of affiliations 
\bibpunct{(}{)}{;}{a}{}{,}
\title[The tilt of the velocity  ellipsoid from RAVE]{Estimation of the
  Tilt of the Stellar Velocity Ellipsoid from RAVE and Implications
  for Mass Models}

\author[A. Siebert et al.]{
A.~Siebert$^{1,2}$\thanks{E-mail: siebert@astro.u-strasbg.fr},
O.~Bienaym\'e$^{1}$,
J.~Binney$^{3}$,
J.~Bland-Hawthorn$^{4}$,
R.~Campbell$^{2,5}$,
\newauthor
K.C.~Freeman$^{6}$,
B.K.~Gibson$^{7}$,
G.~Gilmore$^{8}$,
E.K.~Grebel$^{9}$,
A.~Helmi$^{10}$,
U.~Munari$^{11}$,
\newauthor
J.F.~Navarro$^{12}$,
Q.A.~Parker$^{4,5}$,
G.M.~Seabroke$^{13,8}$,
A.~Siviero$^{11,2}$,
M.~Steinmetz$^{2}$,
\newauthor
M.~Williams$^{2,6}$,
R.F.G.~Wyse$^{14}$,
T.~Zwitter$^{15}$\\
$^1$Universit\'e de Strasbourg, Observatoire  Astronomique,  Strasbourg,  France \\
$^2$Astrophysikalishes  Institut Potsdam  , Potsdam, Germany\\
$^3$Rudolf Peierls Centre for Theoretical Physics, Oxford, UK\\
$^4$Anglo-Australian Observatory, Sydney, Australia \\
$^5$Macquary University, Sydney, Australia\\
$^6$Australian Natianal University, Canberra, Australia\\
$^7$University of Central Lancashire, Preston, UK\\
$^8$Institute of Astronomy, Cambridge, UK\\
$^9$Astronomisches Rechen-Institut, Zentrum f\"ur Astronomie der
Universit\"at Heidelberg, Heidelberg, Germany\\
$^{10}$Kapteyn Astronomical Institut, Groningen, the Netherlands\\
$^{11}$Astronomical Observatory of Padova in Asiago, Asiago, Italy\\
$^{12}$University of Victoria, Victoria, Canada\\
$^{13}$e2v Centre for Electronic Imaging, Planetary \& Space Sciences
Research Institute, The Open University, Milton Keynes, UK\\
$^{14}$Johns Hopkins University, Baltimore, MD, USA\\
$^{15}$Faculty of Mathematics and Physics, University of Ljubljana, Ljubljana, Slovenia
}

\begin{document}

\date{Accepted 2008 September 2. Received 2008 August 28; in original form
  2008 August 7}

\maketitle

\begin{abstract}
We present a  measure of the inclination of the  velocity ellipsoid at 1~kpc
below the  Galactic plane using a sample  of red clump giants  from the RAVE
DR2  release.  We find  that the  velocity ellipsoid  is tilted  towards the
Galactic plane with an inclination of $7.3\pm1.8^{\circ}$.
\noindent We compare this value to computed inclinations for two mass models
of the Milky  Way.  We find that our measurement is  consistent with a short
scale  length of the  stellar disc  ($R_d\simeq2$~kpc) if  the dark  halo is
oblate or  with a long scale  length ($R_d\simeq3$~kpc) if the  dark halo is
prolate.
\noindent Once  combined with independent  constraints on the  flattening of
the halo,  our measurement suggests  that the scale length  is approximately
halfway  between  these  two  extreme   values  ,  with  a  preferred  range
[$2.5$-$2.7$]~kpc for a nearly spherical halo. Nevertheless, no model can be
clearly ruled  out.  With the  continuation of the  RAVE survey, it  will be
possible  to provide a  strong constraint  on the  mass distribution  of the
Milky  Way using  refined measurements  of the  orientation of  the velocity
ellipsoid.
\end{abstract}


%%%%%%%%%%%%%%%%%%%%%%%%%%%%%%%%%%%%%%%%%%%%%%%%%%%%%%%%%%%%%%%%%%%%%%%%%%%
%
%
%%%%%%%%%%%%%%%%%%%%%%%%%%%%%%%%%%%%%%%%%%%%%%%%%%%%%%%%%%%%%%%%%%%%%%%%%%%
\section{Introduction}

Our  understanding  of Galactic  stellar  populations  and kinematics  makes
regular  progress with  the advent  of  new large  Galactic stellar  surveys
providing distances,  photometry, radial velocities or  proper motions.  Our
Galaxy is  at the present  the only  place where we  can probe the  6D phase
space of  stellar positions and  velocities.  For instance, the  Galactic 3D
potential  can  be probed  through  the  orbits  of the  Sagittarius  stream
\citep{ib01,rm05,newberg02,fel06,helmi04}   or   Palomar   5   tidal   tails
\citep{od03,gd06,gj06},   or   through   the   kinematics  of   halo   stars
\citep{ba05}.  At smaller scales, the potential can also be analysed through
the force perpendicular  to the galactic plane (Oort  1960; Cr\'ez\'e et al.
1998; Kuijken \&  Gilmore 1989a,b,c,1991; Siebert et al.   2003; Holmberg \&
Flynn 2004) or through the coupling between the 3 components of the velocity
in the solar neighbourhood \citep{bien99}.\\

Here,  we concentrate on  the question  of the  orientation of  the velocity
ellipsoid that is  known to be tightly related to the  shape and symmetry of
the galactic potential \citep{oll62,hor63,lyn62,ac91}.

In spite of the long interest in this problem, measuring observationally the
orientation  of the  velocity ellipsoid  outside of  the galactic  plane has
proven to be  very difficult. This is due mainly to  the absence of reliable
distances away  from the Solar neighbourhood.  Despite  this limitation, the
first stellar  stream detected within the  Milky Way halo  towards the north
Galactic pole  by \citet{maj96} shows  a velocity tilt, the  ellipsoid being
inclined  towards  the Galactic  plane.  This  tilt  could result  from  the
expected  velocity correlation induced  by a  spheroidal potential  if these
stars had similar  integrals of motion \citep{bien98}. However  we note that
this stream is not detected locally in the RAVE data \citep{seabroke08}.

Building  realistic Galactic  potentials shows  that  the main  axis of  the
velocity  ellipsoid, at  1\,kpc  above  the Galactic  plane,  points in  the
direction of the $z$-axis of symmetry  of the Galaxy towards a point located
at 5 to 8 kpc behind  the Galactic centre: for instance from numerical orbit
computations  \citep{bin83,kg89a} or applying  to the  \citet{ci87} Galactic
potential  the  \citet{ac91}  formulae.   Such  estimates  of  the  velocity
ellipsoid tilt are necessary for an accurate determination of the asymmetric
drift\footnote{The asymmetric drift is the tendency of a population of stars
  to lag behind the local standard  of rest for its rotational velocity, the
  lag increasing  as a  function of age.}   \citep{bt87}, and for  a correct
measurement   of   the   force   perpendicular   to   the   Galactic   plane
\citep{sta89}.\\

In this  paper, we study the  2D velocity distribution  perpendicular to the
Galactic  plane  for a  sample  of  red clump  stars  from  the RAVE  survey
\citep{dr1,dr2}.   These stars  are  selected between  500\,pc and  1500\,pc
below  the Galactic  plane and  provide  a measurement  of the  tilt of  the
velocity ellipsoid at  $\simeq$1~kpc.  In Section~\ref{s:sample}, we present
the  selection  of the  sample  while  Section~\ref{s:tilt}  focuses on  the
measurement   of  the   inclination  and   possible  biases.    Finally,  in
Section~\ref{s:model}  we compare our  measurement to  computed inclinations
for two extreme  classes of mass models and we  discuss possible outcomes of
this measurement.

%%%%%%%%%%%%%%%%%%%%%%%%%%%%%%%%%%%%%%%%%%%%%%%%%%%%%%%%%%%%%%%%%%%%%%%%%%%
%
%
%%%%%%%%%%%%%%%%%%%%%%%%%%%%%%%%%%%%%%%%%%%%%%%%%%%%%%%%%%%%%%%%%%%%%%%%%%%
\section{Selection of the sample}
\label{s:sample}

Our  sample  is drawn  from  the  second data  release  of  the RAVE  survey
\citep{dr2}  containing about  50~000 stellar  radial velocities  and 20~000
measurements of  stellar parameters.  We  focus on red clump  giants towards
the South Galactic pole to maximize the distance from the plane and minimize
the interstellar  extinction. Hence,  we select our  targets in a  cone with
$b<-60^{\circ}$,  and   we  use  a   colour--magnitude  criterion  following
\citet{veltz} to  select our candidate  red clump stars: 2MASS  $J-K$ colour
within 0.5--0.7 and $K<9.3$.

This colour--magnitude  cut selects mainly red clump  stars whose luminosity
function    (LF)    is   well    defined    and   approximately    Gaussian:
$M_\mathrm{K}=-1.6\pm0.03$.   Also,   the  red  clump  LF   is  narrow,  the
dispersion of the Gaussian LF being  0.22 mag in the $K$-band, and is nearly
independent  of the  metallicity  \citep{alv00}.  It  makes this  population
particularly suited  to study  the kinematics of  stars away from  the solar
neighbourhood  as reliable  distance  estimates can  be  obtained. Also  the
extinction  in the  $K$-band remains  low, $<\!\!\!\!A_K\!\!\!\!>=0.007$~mag
with  a maximum  extinction of  $A_K=0.05$~mag for  this region  of  the sky
\citep{schlegel}.   Hence, extinction does  not contribute  significantly to
our error budget: the average error on  the distance is less than 1\% with a
maximum value of $\sim$2\% for the limiting magnitude of our sample.\\

The selection  criterion, retaining  only the objects  with a  proper motion
value  in  the  RAVE  catalogue,  restricts  the sample  to  763  red  clump
candidates spanning a  distance interval from the Sun of  500 to 1500~pc.  A
small fraction of  these selected stars are dwarfs  or subgiants.  According
to  the  photometric  and  kinematic   modelling  of  the  SGP  and  NGP  by
\citet{veltz}, we can estimate that, at the limiting magnitude of our sample
$m_{\rm K}$=9.3,  75\% of the  sample are red  clump stars, 10\%  dwarfs and
15\% subgiants.  Brighter  than this limit, the fraction  of red clump stars
is larger  and the  quoted fractions are  upper limits to  our contamination
fraction.\\

\begin{figure*}
\centering
\includegraphics[width=5cm]{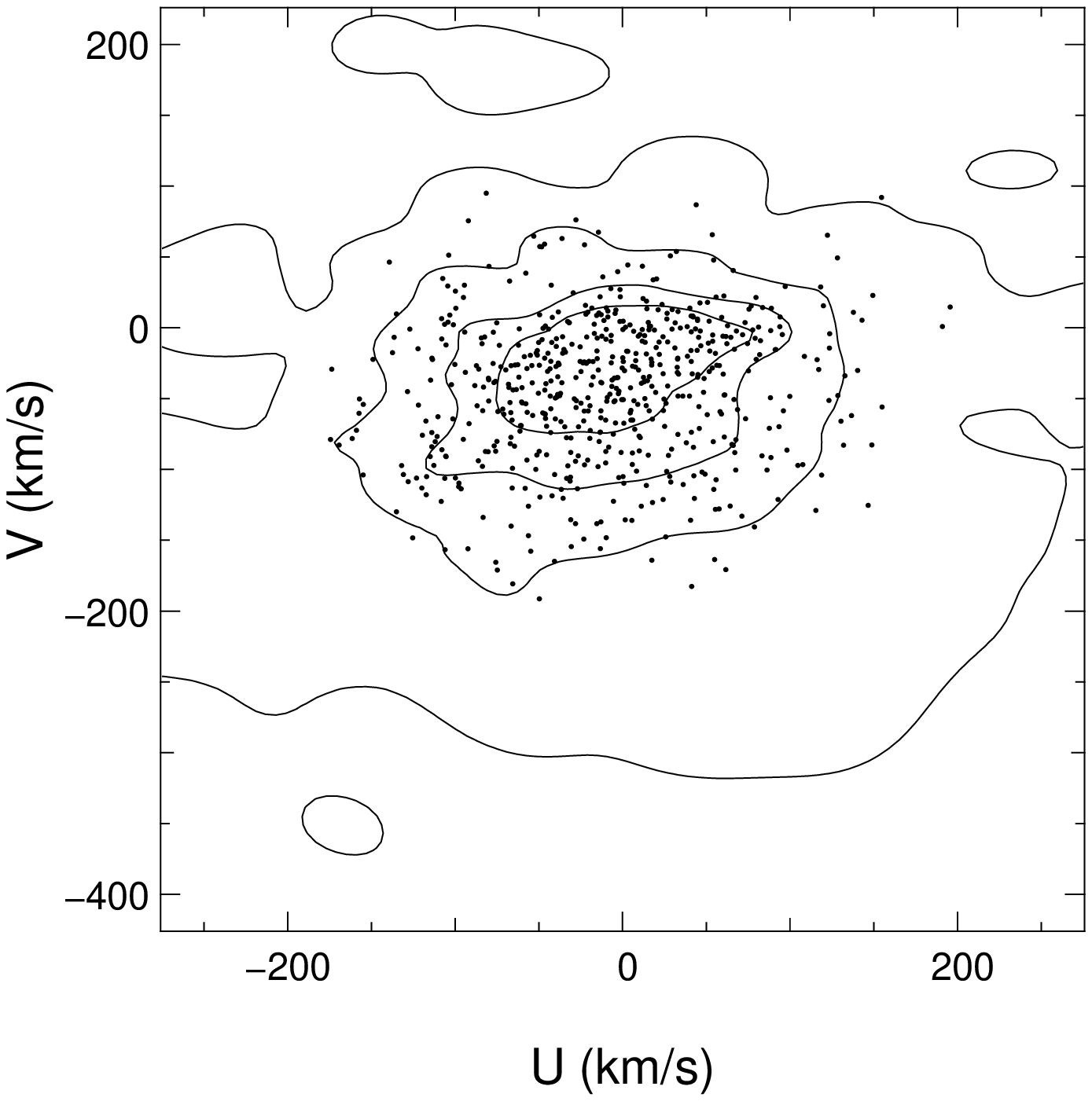}
\includegraphics[width=5cm]{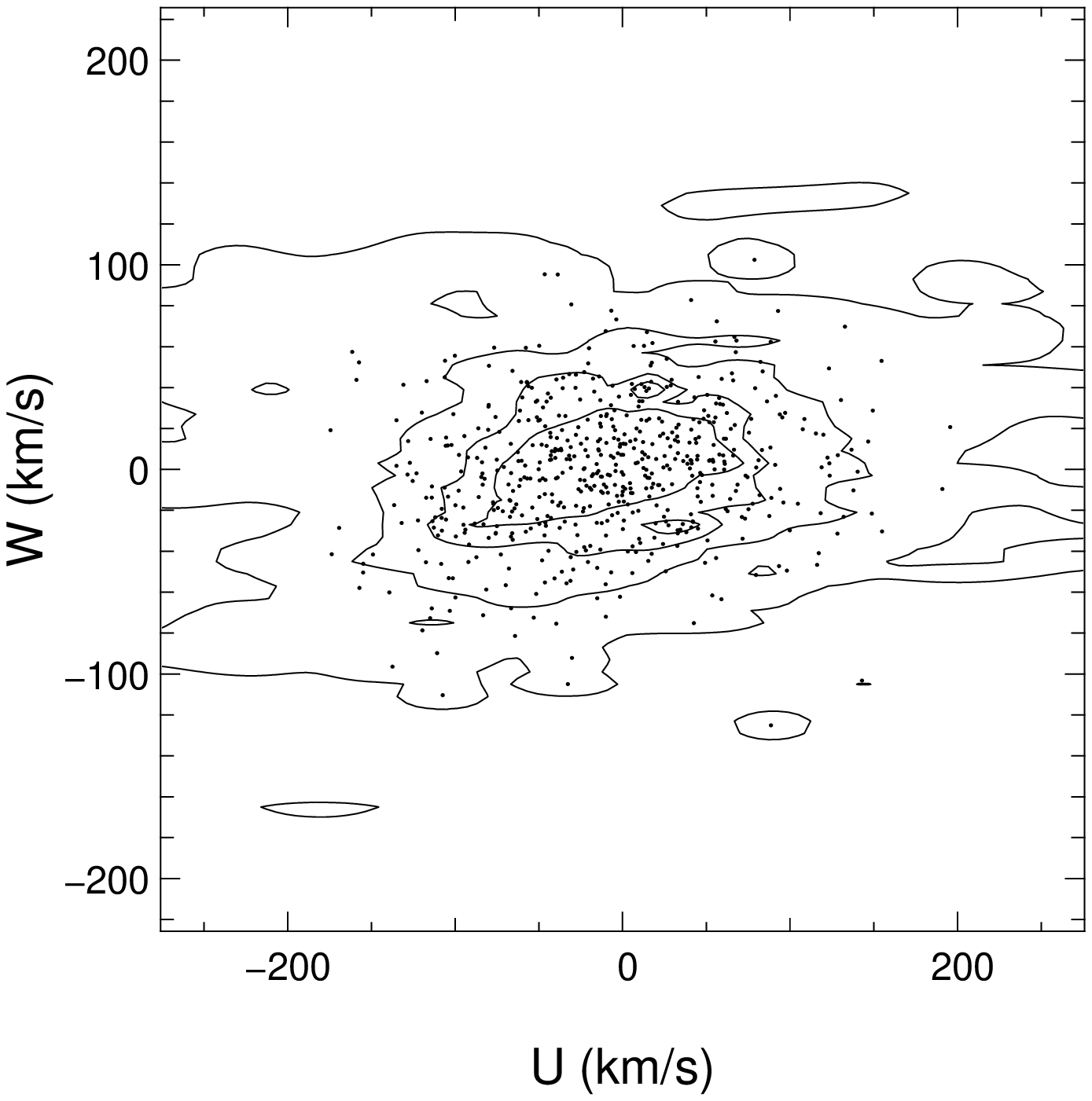}
\includegraphics[width=5cm]{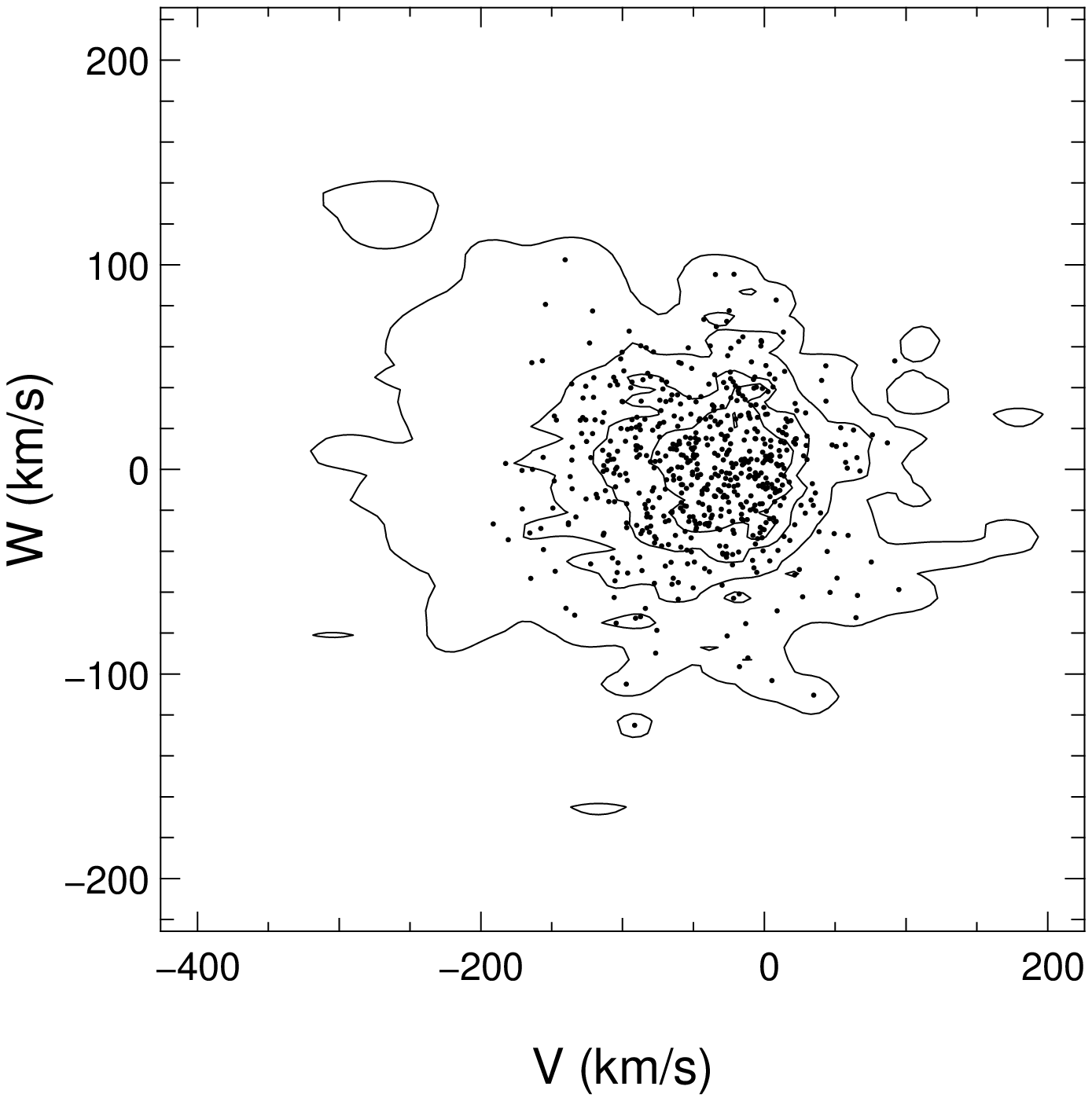}
\caption{Selection of the red clump  sample in $U$, $V$, $W$ velocity space.
  The  contours  show the  distribution  of  the  763 red  clump  candidates
  belonging to the original sample, smoothed by the individual errors, while
  the dots represent the location in  the velocity space of the 580 stars in
  the final  sample.  The  contours encompass 90,  70 ,  50 and 30\%  of the
  total sample.}
\label{f:UVW}
\end{figure*}

We clean  our sample further using  a kinematic selection.   We select stars
based    on     their    velocities    with     the    following    criteria
$\sqrt{U^2+V^2+W^2}<200$~km~s$^{-1}$    and    $V<100$~km~s$^{-1}$.     This
selection  enables  us  to  remove  the  nearby  dwarfs  whose  distance  is
overestimated by a factor 14  due to their fainter absolute magnitude, hence
an overestimation of their velocities. The resulting sample contains 580 red
clump  candidates  in  the  direction  of  the  South  Galactic  pole  whose
distribution in  velocity space is  presented in Fig.~\ref{f:UVW}.   In this
figure, the contours depict the distribution of the original sample smoothed
by the individual errors while the  dots show the location in velocity space
of the remaining 580 stars after the velocity selection.\\

\begin{figure}
\centering
\includegraphics[width=7cm]{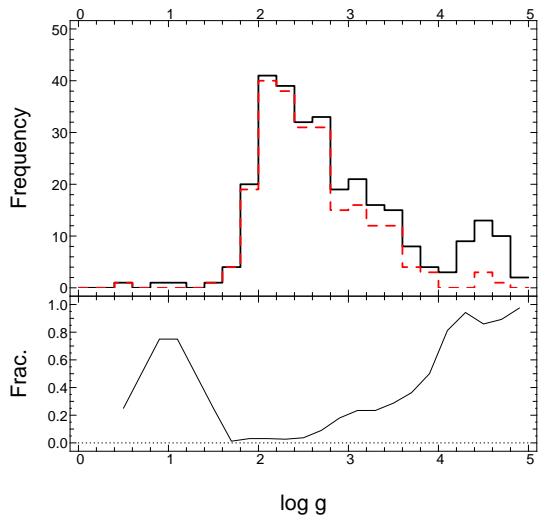}
\caption{Top  panel: histogram  $\log g$  for the  294 stars  with  $\log g$
  measurements in RAVE DR2. Black line full subsample, dashed line subsample
  matching the velocity  criteria. The red clump giants  cover the region in
  $\log g=2-3$ depending on the metallicity or mass. A conservative estimate
  of RAVE standard error on $\log  g$ is 0.5~dex.  Bottom panel: fraction of
  stars rejected by the velocity criterion as a function of $\log g$. }
\label{f:logg}
\end{figure}

We test  our selection criteria using  the second year  observation from the
RAVE survey (as a reminder, RAVE  DR2 contains the first year of observation
--i.e. DR1-- and  the second year of observation).   For these objects, RAVE
provides measurements of the stellar parameters including an estimate of the
gravity.  The sample selected from  second year data contains 294 stars with
$\log g$ measurements,  with 231 stars matching the  velocity criteria.  The
histograms of $\log g$ for each subsample are presented in Fig.~\ref{f:logg}
where the black histogram presents the  distribution of $\log g$ for the 294
second  year stars  and  the  dashed histogram  the  subsample matching  our
velocity  criteria. The  red  clump giants  span  a large  range in  gravity
depending on their metallicity: $\log  g=2.08$ for the metal--poor, low mass
end and  reaches up to $\log g=3$  for the high mass,  metal--rich red clump
objects  \citep{za00}.   This figure  clearly  indicates  that our  velocity
criteria is  efficient for  rejecting dwarf stars  (with high $\log  g$) but
also  removes a  small  fraction of  stars  with lower  $\log g$,  primarily
subgiants and giants on the ascending branch and also a few red clump stars.
Nevertheless, these objects  have large velocities and fall  in the tails of
the  velocity  distribution.  Therefore,  they  affect  only marginally  the
measurement of the  inclination, as our measurement is  driven by the larger
number of stars in the bulk of the velocity distribution.\\

It  is  worth  noting  that  due  to the  uncertainties  in  RAVE  $\log  g$
measurements which  are 0.5~dex for a  typical RAVE star  \citep{dr2}, it is
not possible to  obtain a firm estimate of the  contamination in our sample,
nor to use the RAVE $\log g$ estimates to refine our sample.  Also, $\log g$
measurements are only available for less  than half of our sample as stellar
parameters can not be estimated  from the spectra collected during the first
year of  operation of  RAVE.  Nevertheless, considering  a 0.5~dex  error on
$\log g$, we estimate that  the contamination using the velocity criteria is
reduced to  $\simeq$10\% which is to  be compared to more  than 20\% without
the velocity criteria.   We will detail the effect  of this contamination on
our measurement in the next section.

%%%%%%%%%%%%%%%%%%%%%%%%%%%%%%%%%%%%%%%%%%%%%%%%%%%%%%%%%%%%%%%%%%%%%%%%%%%
%
%
%%%%%%%%%%%%%%%%%%%%%%%%%%%%%%%%%%%%%%%%%%%%%%%%%%%%%%%%%%%%%%%%%%%%%%%%%%%
\section{Measuring the tilt}
\label{s:tilt}

The tilt  angle $\delta$  of the  2D velocity distribution  is given  by the
relation:

\begin{equation}
\tan 2\delta    =    \frac{2\sigma_{UW}^2}{\sigma_{U}^2-\sigma_{W}^2}\,,
\label{e:tilt}
\end{equation}

\noindent where  $\sigma_{UW}^2$, $\sigma_{U}^2$ and  $\sigma_{W}^2$ are the
velocity  distribution  moments.  In  the  local  velocity coordinates,  the
velocity  in the  radial direction  is  given by  the $U$  component of  the
velocity  vector (positive  towards the  Galactic centre)  and  the vertical
velocity  by the  $W$ component  of the  vector positive  towards  the North
Galactic  pole, while  the $V$  component is  positive towards  the Galactic
rotation (not used in Eq.~\ref{e:tilt}).\\

The  computation of  the inclination  is straightforward  for a  sample with
small  and  homogeneous  errors.   Nevertheless,  to  lower  the  effect  of
foreground  dwarfs and  giants and  the contamination  due to  high-velocity
stars, we  make use of  a velocity cut--off  to select our sample.   In this
case, as  our errors  in the $U$  and $V$  velocity directions are  large, a
direct  measurement  of the  tilt  angle  may be  subject  to  bias and  our
selection criteria must be studied as  our error budget may not be dominated
by the size of the sample (see Section~\ref{s:anisotropy}).

Also, the local  velocity ellipsoid is not a  smooth distribution and clumps
are present on both small and  large scales in the velocity space \citep[see
  for  example][]{deh00,deh98b,che98,fam05}.   These  substructures  prevent
determining  the  age-velocity  dispersion  relation  in  the  $U$  and  $V$
directions \citep{sg07} and  may also influence the measured  tilt angle. We
will  discuss  the  effect  of  such substructures  on  our  measurement  in
Section~\ref{s:substructure}.

Finally, if our selection criterion is efficient at rejecting the foreground
stars,  only the  tails of  the velocity  distribution are  affected  by the
velocity  cut--off.   As  their  space velocities  are  overestimated,  such
foreground objects  will impact  on the measurement  of the  inclination. We
will discuss this particular point in Section~\ref{s:foreground}.

%%%%%%%%%%%%%%%%%%%%%%%%%%%%%%%%%%%%%%%%%%%%%%%%%%%%%%%%%%%%%%%%%%%%%%%%%%%
\subsection{The effect of errors and velocity cut--off}
\label{s:anisotropy}

\begin{figure}
\includegraphics[width=7cm]{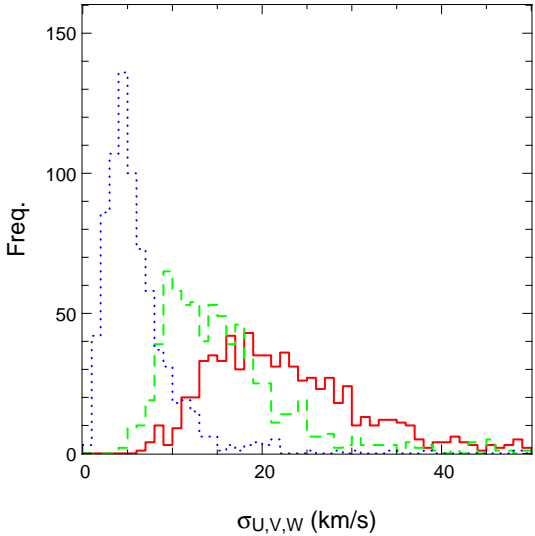}
\caption{Distribution of the  velocity errors in our sample;  full line $U$
  component,  dashed line  $V$ component,  dotted line  $W$  component.  The
  difference  between this three  distributions arises  due to  the relative
  contribution of proper  motion and distance errors to  the radial velocity
  errors for the velocities along the three cardinal directions.}
\label{f:vel_error}
\end{figure}

Our velocity errors  in the cardinal directions are  not homogeneous because
the  $U$ and  $V$ components  of the  velocity vector  are dominated  by the
proper  motion contribution while  the $W$  component is  primarily measured
from   the  RAVE  radial   velocity.  Fig.~\ref{f:vel_error}   presents  the
distribution of errors for our sample  in the $U$, $V$ and $W$ components as
full, dashed  and dotted lines.  It is  clear that the mode  of the velocity
error distributions for the $U$ and $W$ components, the ones we are primarly
interested  in, differ  by a  factor 4:  $\simeq$~5~km~s$^{-1}$ for  the $W$
component   while  for  the   $U$  component   the  distribution   peaks  at
$\simeq$~20~km~s$^{-1}$.

This large  difference results in  an anisotropic smoothing of  the observed
velocity ellipsoid  which, combined with our velocity  criterion, biases the
measurement of  the inclination towards a  lower value. This bias  is due to
the structure of  Eq.~\ref{e:tilt} where the error anisotropy  results in an
extra component  $E$ on measured  velocity dispersions. If we  consider only
the     extra     term     on      the     $U$     component     we     have
$\sigma_U(measured)^2=\sigma_U(true)^2+E^2$      and      the     cross-term
$\sigma_{UW}(measured)=\rho_{UW}\sigma_W\sqrt{\sigma_U(true)^2+E^2}$,
$\rho_{UW}$  being the correlation  coefficient. With  this notation,  it is
clear that the  contribution of the additional error term  is larger for the
denominator than it  is for the numerator, hence  producing an underestimate
of the  true inclination. For comparison,  a linear fit would  not be biased
due to the asymmetry of the errors but unfortunately it is more sensitive to
outliers.  To  overcome this problem, we  use a Monte Carlo  sampling of the
velocity error distributions.  We add a  random velocity term to the $V$ and
$W$ components,  degrading the accuracy  of the two velocity  components, so
that  the  resulting  error  distributions  match  the  $U$  velocity  error
distribution. For the  $U$ velocity, it is randomly  drawn from its original
error  distribution. This  procedure enables  us to  obtain  isotropic error
distributions for all three components, degrading the two best distributions
to the  level of  the least accurate  distribution. The inclination  is then
computed using Eq.~\ref{e:tilt} after applying the velocity criterion.

\begin{figure}
\includegraphics[width=7cm]{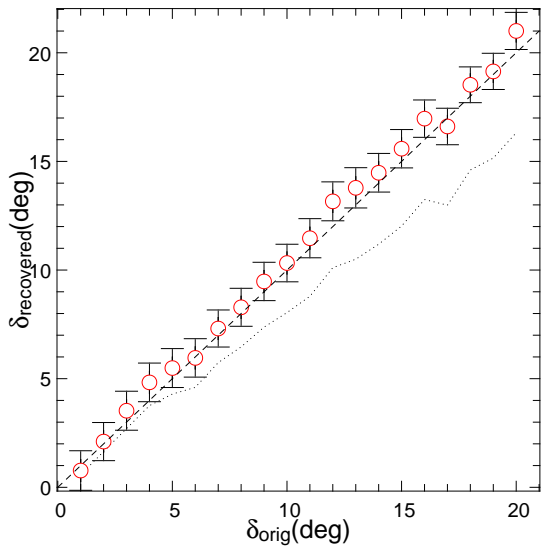}
\caption{Test of  the Monte Carlo  method to measure  the tilt angle  of the
  velocity ellipsoid. The original versus recovered tilt angle are presented
  for  one realisation  of  a velocity  ellipsoid  having $\sigma_U=31$  and
  $\sigma_W=17$~km~s$^{-1}$ and  sampled using  1000 data points.   The RAVE
  error laws for  the velocities are used. The dashed line  shows the one to
  one  relation  while  the  dotted  line  is  a  direct  measurement  using
  Eq.~\ref{e:tilt}. The results  from the Monte Carlo sampling  of the error
  laws are depicted by the open  circles and the error bars are the standard
  deviation of 5000 resampling for each value of the tilt angle.}
\label{f:test}
\end{figure}

We tested this procedure on a simple velocity ellipsoid model using the RAVE
error  laws and  standard velocity  dispersions  for the  Galactic old  disc
population,    leaving   aside   the    thick   disc:    $\sigma_U=31$   and
$\sigma_W=17$~km~s$^{-1}$.   The size  of the  sample was  set to  1000 data
points  and  we   varied  the  inclination  of  the   ellipsoid  from  1  to
20$^{\circ}$.  The results of this test are shown in Fig.~\ref{f:test} where
the direct  measurement is presented  as a dotted  line and the  Monte Carlo
determinations  by  the  open  circles   with  error  bars  for  one  random
realisation of  a velocity ellipsoid.  The  one to one  relation between the
original  and  recovered angles  is  sketched  by  the dashed  line.   Below
2-4$^{\circ}$  for the  inclination,  depending on  the  realisation of  the
ellipsoid,  both  methods  predict  the  same  inclination  but  above  this
threshold,  the  Monte Carlo  sampling  recovers  the  proper value  of  the
angle. On the other hand, the direct measurement, applying Eq.~\ref{e:tilt},
always underestimates the true angle with a bias rising with the tilt value.
This test clearly  indicates that the Monte Carlo sampling  of the errors is
best  suited to measure  the tilt  of the  velocity ellipsoid,  while direct
measurements using Eq.~\ref{e:tilt}  are subject to strong bias  in the case
of  heterogeneous error  laws.  We  note  also that  the value  of the  bias
depends  strongly on  the  random sampling  of  the ellipsoid,  with a  bias
varying between 2  and 4$^{\circ}$ at 7$^{\circ}$.  This  spread in the bias
value becomes  larger as  the tilt value  increases and indicates  that even
with a proper model  to estimate the bias, it can hardly  be used to correct
the direct measurement.

\begin{figure}
\centering
\includegraphics[width=7cm]{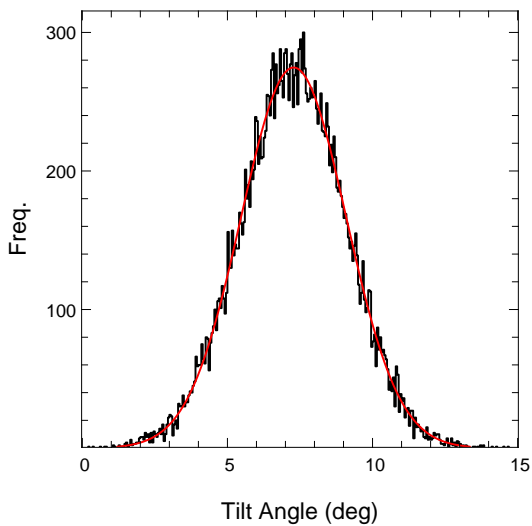}
\caption{Distribution of the measured  inclination of the velocity ellipsoid
  per 0.05$^{\circ}$ bin.  This distribution is obtained using a Monte Carlo
  sampling of  the error distribution. The  mean inclination is  found to be
  7.3$^{\circ}$ with a standard deviation of 1.8$^{\circ}$. The grey line is
  a Gaussian function with identical parameters.}
\label{f:monte-carlo}
\end{figure}

\begin{figure}
\centering
\includegraphics[width=7cm]{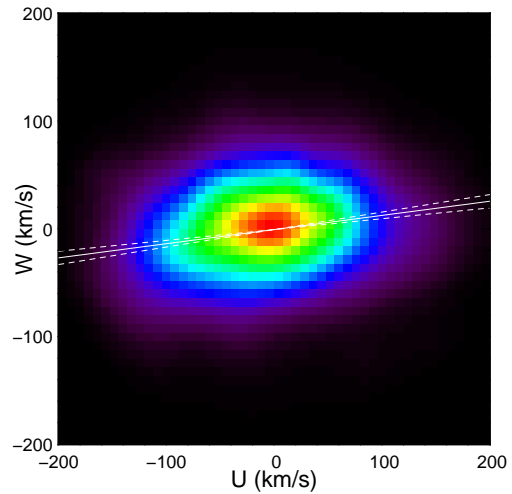}
\caption{Velocity distribution in the  ($U$,$W$) plane from our sample after
  sampling the  error distribution. The measured  inclination and 1-$\sigma$
  range are presented by the  full and dashed white line. The colour--coding
  follows the density per bin.}
\label{f:tiltfinal}
\end{figure}

The procedure is then applied on the RAVE sample and the result is presented
in Fig.~\ref{f:monte-carlo} which shows  the distribution of inclinations in
degrees obtained by sampling the  error distribution 25~000 times.  The mean
inclination  measured   is  7.3$^{\circ}$  with  a   standard  deviation  of
1.8$^{\circ}$.  If the anisotropy of  the error distributions were not taken
into  account, the  measured inclination  would have  been  6.3$^{\circ}$ or
1.0$^{\circ}$  too low.   The 2D  representation of  the  velocity ellipsoid
inclination is shown in Figure~\ref{f:tiltfinal}. The colour--coding follows
the density of stars per bin in the region of the ($U$,$W$) space, where the
2D  distribution of  the  ($U$,$W$)  velocities has  been  convolved by  the
individual  errors.   The measured  inclination  and  1-$\sigma$ errors  are
presented as white lines (full line  for the mean value and dotted lines for
the errors).

%%%%%%%%%%%%%%%%%%%%%%%%%%%%%%%%%%%%%%%%%%%%%%%%%%%%%%%%%%%%%%%%%%%%%%%%%%%
\subsection{Effect of substructures}
\label{s:substructure}

Substructures such as  the Hyades, Pleiades or the  Hercules groups are well
known features of  the local velocity ellipsoid, and are  easily seen in the
velocity  space obtained  from the  {\it Hipparcos}  mission  \citep[see for
  example][]{fam05,deh98,che98}.  These structures may  have a wide range of
origins such as  the disruption of clusters, resonances  associated with the
bar or spiral arms \citep[see for example][]{fam05,fam08,min08,deh00,des04}.
Nevertheless, the average velocity error in  our sample does not allow us to
distinguish these substructures.\\

To  test  the  influence  of   these  velocity  substructures  on  the  tilt
determination,  we use  the local  sample from  \citet{fam05}.   This sample
provides  not only accurate  velocity vectors  for about  6500 stars  in the
solar neighbourhood, it also provides an  estimate of the relation of a star
to the  identified velocity  substructures. This allows  us to  separate the
background ellipsoid from the known overdensities.\\

We  first   estimate  the  fraction   of  stars  in  substructures   in  the
\citet{fam05} sample  as a  function of $z$,  the height above  the Galactic
plane. The number of stars in structures is larger closer to the plane: 36\%
of the  stars are  in structures  in the 0--200~pc  interval while  25\% are
found in structures between 200 and  500~pc.  This drop in number of objects
in structures  is sharp as  seen from Table~\ref{t:z_frac}, the  fraction in
objects in structures being lowered by over a factor 2 between 0 and 500~pc.
As our  sample covers  a distance below  the plane  from 500 to  1500~pc, we
extrapolate  this behaviour  at  higher $z$  to  estimate the  contamination
arising   from  substructures   in   our  sample.    Using  a   conservative
extrapolation, we estimate the contamination  in our sample to be lower than
7\%.\\

\begin{table}
\caption{Fraction  and number of  stars in  structures in  the \citet{fam05}
  sample as a function of height above the Galactic plane.}
\label{t:z_frac}
\centering
\begin{tabular}{@{}cccccc}
\hline
z    & N$_{tot}$ & Fraction in & Num. in  & Num. in & Num. in\\
(pc) &           & structures  & Hya/Plei & Sirius  & Hercules\\
\hline
  0-100 & 1361 & 0.40 & 127 & 63 & 125\\
100-200 & 1337 & 0.33 & 111 & 49 & 131\\
200-300 &  844 & 0.28 & 46  & 27 & 94\\
300-400 &  422 & 0.22 & 1   & 20 & 37\\
400-500 &  177 & 0.19 & 0   & 7  & 8\\
\hline
\end{tabular}
\end{table}

In a second step, we test the influence of the substructures on the measured
inclination.   We can  not use  the  \citet{fam05} sample  directly, as  the
presence of pertubations  in the plane makes any  attempt to disentangle the
effect  of groups from  the effect  of the  pertubations on  the inclination
hazardous. Therefore, we proceed as follow. We add an additional population,
drawn from a subset of the Famaey  et al.  sample, to our RAVE sample.  This
subset is randomly selected from the set of stars belonging to groups in the
distance interval 300 to 500\,pc.  We  add it to the RAVE sample varying its
fraction relative to the RAVE sample  from 1 to 20\%. This procedure enables
us to mimic as closely as  possible the velocity distribution of the groups,
which is  not homogeneous and strongly  varies as a function  of distance to
the plane.  This operation is repeated 25~000 times for each fraction of the
contamination to ensure a proper coverage of the possible cases.\\

\begin{figure}
\centering
\includegraphics[width=7cm]{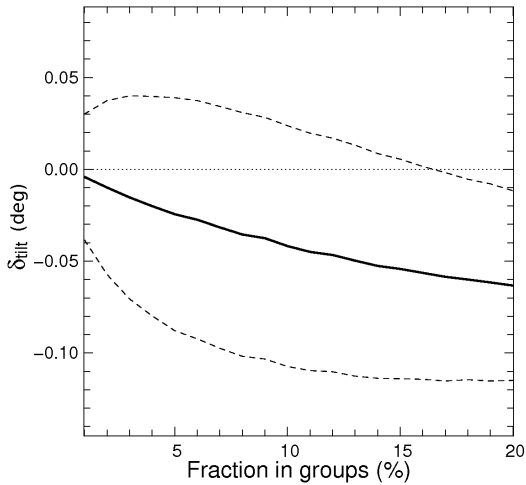}
\caption{Influence of stars in  velocity groups on the measured inclination.
  A sample of  \citet{fam05} stars belonging to groups  is randomly added to
  the RAVE  sample, varying the  contamination from 1  to 20\%. The  tilt is
  measured following  the same  procedure as for  the pure RAVE  sample. The
  thick    line    represents    the    average   deviation    in    degrees
  ($\delta_{tilt}=\delta_{measured}-\delta_{true}$) found for 25~000 repeats
  per contamination fraction.  The dashed  line is the standard deviation of
  the repeats.}
\label{f:struct_influence}
\end{figure}

The  results  are  shown  in Fig.~\ref{f:struct_influence}  where  the  mean
deviation  ($\delta_{measured}-\delta_{true}$) in degrees  as a  function of
the fraction  of stars  in groups is  drawn as  a thick line.   The standard
deviation of the repeats is shown as dashed lines.  We note that the average
tilt for  the Famaey group  sample used here is  $0.07\pm2.45^{\circ}$, very
different from the value of the inclination in our sample. From this figure,
we see  that the influence  of velocity structures  on the measured  tilt is
low,  around $-0.03^{\circ}$  for a  contamination  of 7\%  with a  standard
deviation  below   $0.1^{\circ}$.   7\%  being   an  upper  limit   for  the
contamination  in  our  sample,  we  do  not expect  groups  to  affect  our
measurement  of the  tilt.  Indeed,  the  mean deviation  combined with  the
standard  deviation measured from  this experiment  contributes to  not more
than $0.1^{\circ}$, less than 6\% of our estimated errors.

\subsection{Effect of foreground stars}
\label{s:foreground}

If, as  we saw above, the  velocity structures do not  influence largely the
orientation  of  the  velocity  ellipsoid,  the  foreground  stars  (dwarfs,
subgiants and giants on the  ascending branch) can be more problematic.  The
fact that their  velocities are overestimated in the  $U$ direction, because
of  the  overstimate of  their  distances, will  add  a  component with  low
inclination to the observed ellipsoid.  The  bias due to these objects is an
underestimate of the tilt at a given distance.\\

The contamination  by foreground  objects is about  10\% in our  sample (see
Section~\ref{s:sample}), and a factor  14 overestimation of the $U$ velocity
for the  dwarfs will render their  velocity ellipsoid almost  uniform in the
velocity interval  we consider. The  velocity dispersion of  this foreground
population    will    be   large,    over    400~km.s$^{-1}$   instead    of
$\sim$31~km.s$^{-1}$ for $\sigma_U$, due  to the distance overestimate.  For
the other  sources of contamination  (subgiants and giants on  the ascending
branch), the  distance is  overestimated by  a factor 2  or less,  and their
impact on the velocity ellipsoid is lower. \\

To obtain an upper limit of  the effect of the foreground population we rely
on  a  resampling  technique, replacing  10\%  of  the  sample by  a  random
realization of a thin disc  population with no inclination. The overestimate
of the distance  is then translated into an  overestimate of the velocities,
and the final inclination of the velocity ellipsoid is measured applying the
same procedure as  above.  The difference between the  distribution with and
without  resampling provides  an upper  limit  on the  effect of  foreground
objects on our  measurement. We note here that, if  adding a population with
no   tilt   is   in   principle   similar  to   the   experiment   done   in
Section~\ref{s:substructure},    here    we    incorporate   the    distance
overestimation. Furthermore, the added test  stars are not restricted to the
region in velocity space of the groups.\\

\begin{figure}
\centering
\includegraphics[width=7cm]{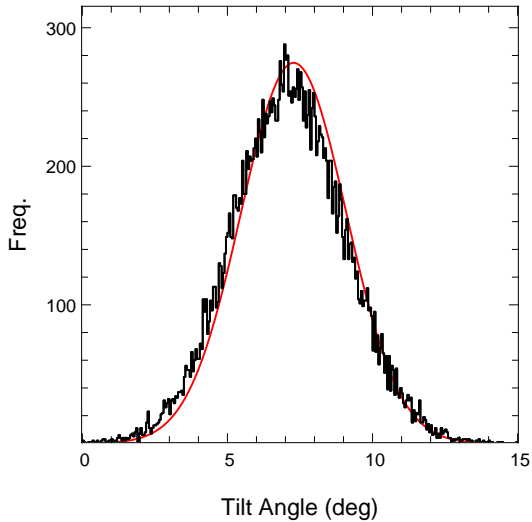}
\caption{Results from the resampling study. 10\% of the red clump sample has
  been  replaced by  a population  of foreground  objects with  no  tilt and
  velocities in the  $U$ direction overestimated by a  factor 14.  The black
  histogram is the distribution of  measured inclination for this new sample
  following  the same  procedure as  for Fig.~\ref{f:monte-carlo}  while the
  grey  Gaussian  curve  is  the  distribution of  inclination  without  the
  resampling from Fig.~\ref{f:monte-carlo}.}
\label{f:resampling}
\end{figure}

Figure~\ref{f:resampling}  presents  the results  of  this resampling.   The
black histogram shows the distribution  of the 25~000 measurements while the
grey  Gaussian curve  is  the Gaussian  representation  of the  distribution
obtained in  Fig.~\ref{f:monte-carlo}. The presence of a  population with no
inclination does produce an observable bias in this experiment: we observe a
shift in the maximum of  the distribution. Nevertheless, this bias is small,
the  measured  offset  is  0.15$^{\circ}$,  much  lower  than  the  standard
deviation  of the  distribution while  it  is a  {\it worst-case  scenario}.
Indeed,  here we  did consider  only the  contamination by  foreground dwarf
stars while our real contamination is a mixture of dwarfs and subgiants.  In
the latter  case, the overestimate of  the distance is much  lower, as these
stars have  a mean  distance from  the plane that  is larger.   Hence, their
impact  on  the  velocity  ellipsoid  orientation will  be  lower  than  for
dwarfs. We  note also that the  presence of a  foreground population renders
the distribution non-Gaussian, adding a tail to the low  tilt angle part of
the distribution which is not observed in Fig.~\ref{f:monte-carlo}.\\

We  can  conclude that  our  estimate of  the  inclination  of the  velocity
ellipsoid  is robust and  that the  presence of  a population  of foreground
objects does not  introduce a significant bias in  our measurement.  Indeed,
combining both the  effect of foreground stars and  of the possible velocity
ellipsoid substructures in a  {\it worst--case} scenario, the resulting bias
amounts  to $\sim$10\%  of our  errors.  At  this level,  the biases  do not
affect our conclusions.

%%%%%%%%%%%%%%%%%%%%%%%%%%%%%%%%%%%%%%%%%%%%%%%%%%%%%%%%%%%%%%%%%%%%%%%%%%%
%
%
%%%%%%%%%%%%%%%%%%%%%%%%%%%%%%%%%%%%%%%%%%%%%%%%%%%%%%%%%%%%%%%%%%%%%%%%%%%
\section{Relation to the mass distribution in the Galaxy}
\label{s:model}

The  tilt  of  the velocity  ellipsoid  is  intimately  linked to  the  mass
distribution  in the  Milky Way  and more  specifically --  if we  trust our
knowledge of the structure of the  Galactic disc -- to the flattening of the
halo. We  start from the  mass model of  \citet{deh98}\footnote{The Galactic
  potentials are  computed using the  GalPot program written by  W.  Dehnen.
  This   program    is   available   within   the    {\sc   NEMO}   package:
  http://carma.astro.umd.edu/nemo/.}  and its revised parameters provided by
\citet{bt08} in  their table~2.3  (hereafter BT08).  This  revision proposes
two  models, referred  to as  model I  and II,  which match  both  local and
non-local  data.  These models  are modified  versions of  the \citet{deh98}
models 1 and 4.

As noted in these references, a crucial parameter is the scale length of the
disc whose value lies  in the range 2-3~kpc.  The two models  are set on the
upper and  lower bound for  this parameter.  Model  I presents a  mass model
with a short scale length  ($R_d=$~2~kpc) that induces a strong contribution
from the disc  for the potential at  the solar radius up to  11~kpc.  On the
other  hand,  model  II  has  a larger  scale  length  ($R_d=$~3.2~kpc)  and
therefore, the halo contribution to  the rotation curve dominates at the Sun
location  and beyond.   This  is also  seen  from the  global  shape of  the
potential where for  model II the isopotentials are  more spherical than for
model I  (see figs.~2.19 and 2.21  of BT08).  For a  detailed description of
these two models, the reader is referred to chapter~2 of BT08.\\

We use  both models  to discuss below  the implications  of the tilt  on the
possible models for the mass distribution  in the Milky Way, focusing on the
flattening  of the halo  in the  two extreme  cases. We  note here  that the
region  above (below)  the  plane between  1 and  2  kpc is  best suited  to
separate the two  classes of models.  Indeed, in  this region, the variation
of the angle between the Galactic  plane and the normal to the isopotentials
as  a function  of the  minor--to--major axis  ratio $c/a_\rho$  is maximum.
Hence, we expect  the difference between the predicted tilt  angle to be the
largest  in  the same  region.   At larger  distances  from  the plane,  the
potential  becomes more spherical  and the  difference vanishes  between the
models  in  terms of  variation  of the  potential  and  inclination of  the
velocity ellipsoid.\\

To measure the  tilt of the ellipsoid as a function  of the halo flattening,
we vary the density minor--to--major  axis ratio $c/a_\rho$ of the halo from
0.6 to 1.7 for each model, the halo density being described by the relation
\begin{equation}
\rho(R,z)=\rho_{halo}\left(\frac{m}{a_h}\right)^{-\alpha_h}   \left(   1   +
\frac{m}{\alpha_h} \right)^{(\alpha_h-\beta_h)},
\end{equation}
where the  flattening $c/a_\rho$ enters  the equation through  the parameter
$m=\sqrt{R^2+z^2/(c/a_\rho)^2}$,  $R$  and   $z$  being  the  Galactocentric
cylindrical  coordinates.  $a_h$  is  a scale  parameter  with  $\rho\propto
m^{-\alpha_h}$ if $m \ll a_h$ and $\rho \propto m^{-\beta_h}$ for large $m$.

While changing  $c/a_\rho$, we change  the halo
density in order  to keep the rotation curve almost  unchanged in the plane.
The bulge and disc components are  fixed to the values of table~2.3 of BT08,
the solar Galactocentric radius$=8$~kpc.  Keeping the rotation curve and the
disc/bulge parameters  unchanged enables  us to study  the influence  of the
halo flattening on the shape  of the velocity ellipsoid, as the contribution
to the  radial force  of each  component remains largely  the same  for each
case.  The corresponding density of the halo for each mass model is reported
in Table~\ref{t:massmodel}.

\begin{table}
\caption{Modification to the mass models I  and II of BT08 table 2.3 used in
  Section~\ref{s:model}.    The  halo   density  is   given   in  $M_{\odot}
  \mathrm{pc}^{-3}$.   The modified  models  are built,  modifying the  halo
  parameters, to keep the rotation  curve almost unchanged in the disc.  The
  bulge and disc parameters are fixed to the BT08 values.}
\label{t:massmodel}
\centering
\begin{tabular}{@{}ccc}
\hline
& \multicolumn{2}{c}{$\rho_{halo}$}\\
\cline{2-3}
$c/a$ &  Model I  & Model II\\
\hline
0.6 & 0.838 & 0.327 \\
0.7 & 0.765 & 0.293 \\
0.8 & 0.711 & 0.266 \\
0.9 & 0.670 & 0.245 \\
1.0 & 0.635 & 0.229 \\
1.1 & 0.608 & 0.215 \\
1.2 & 0.585 & 0.204 \\
1.3 & 0.566 & 0.195 \\
1.4 & 0.548 & 0.186 \\
1.5 & 0.534 & 0.179 \\
1.6 & 0.520 & 0.172 \\
1.7 & 0.510 & 0.167 \\
\hline
\end{tabular}
\end{table}

The  inclination of  the velocity  ellipsoid is  computed for  a  given mass
distribution using orbit  integration. A single orbit is  integrated over 30
rotations using a 4th  order Runge--Kutta algorithm.  The initial conditions
are  drawn  from  a  Shu  distribution  function  matching  the  local  data
\citep{bien99}.   For each  potential,  the orbit  library  contains over  2
million orbits from which we randomly select 10 points per orbit in the last
15 rotations.  We further restict  the orbit library to data points matching
the interval in $R$ and $z$ of the RAVE sample, this reduces the size of our
final libraries to $6.10^4$ to a few $10^5$ points per library.

We measure the  tilt using a Monte Carlo selection  of the orbits, requiring
that the distribution of the  selected orbits matches the selection function
of our  RAVE sample in the  ($R$,$z$) plane. Here $R$  is the Galactocentric
radius  and  $z$ the  distance  above  (below)  the plane.   This  selection
function is  obtained by convolving the  distribution of the  RAVE sample in
$R$ and  $z$ by their errors for  each star. This procedure  ensures us that
the spatial distribution  in $R$ and $z$ of the RAVE  sample is well matched
by  the  orbit  selection.   We   select  5~000  orbits  using  the  spatial
constraints, about 10 times larger than the observed sample but about $10^2$
times less than  orbit library size to minimize the  probability of the same
orbit to  be selected twice, and  the tilt is measured  using the associated
velocities and  Eq.~\ref{e:tilt}.  The measurement is repeated  500 times to
obtain the mean inclination and  dispersion.  This procedure is repeated for
each  orbit library.   The convergence  of  this procedure  is tested  using
1~000, 5~000 and 10~000 orbits points  for the selection. The result shows a
very good stability of the mean inclination, of the order of a few $10^{-2}$
degrees,  while the  dispersion increases  as the  number of  orbits becomes
lower.  For  the model  I with $c/a_{\rho}=1.0$,  we obtain  respectively an
inclination  of $9.89\pm1.09$,  $9.83\pm0.44$ and  $9.85\pm0.36$  for 1~000,
5~000 and  10~000 orbits, which indicates  that the gain  in precision above
5~000  orbits is  limited as  the computing  time scales  linearly  with the
number of orbits.

\begin{figure*}
\centering
\includegraphics[width=7cm]{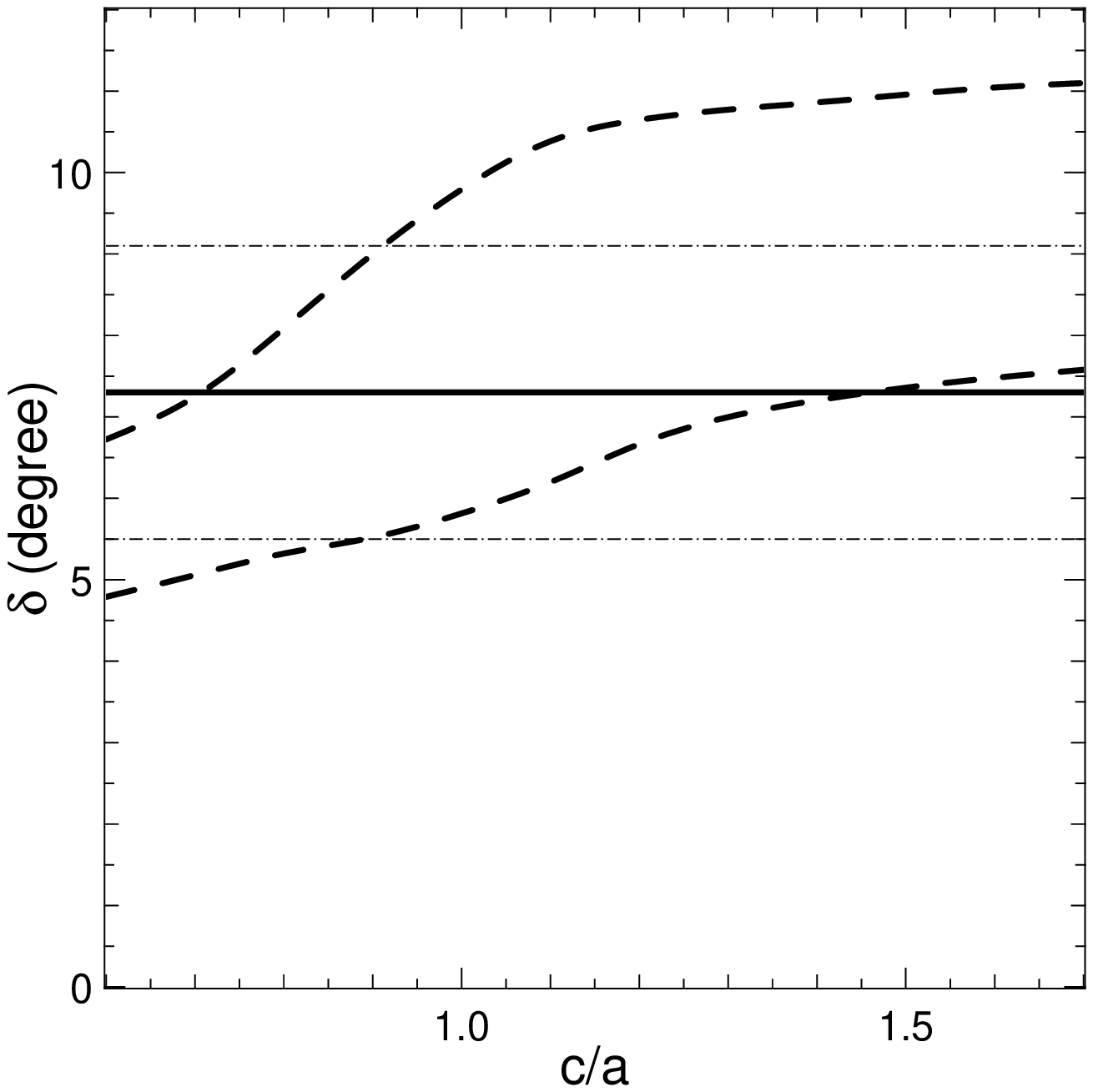}
\includegraphics[width=7cm]{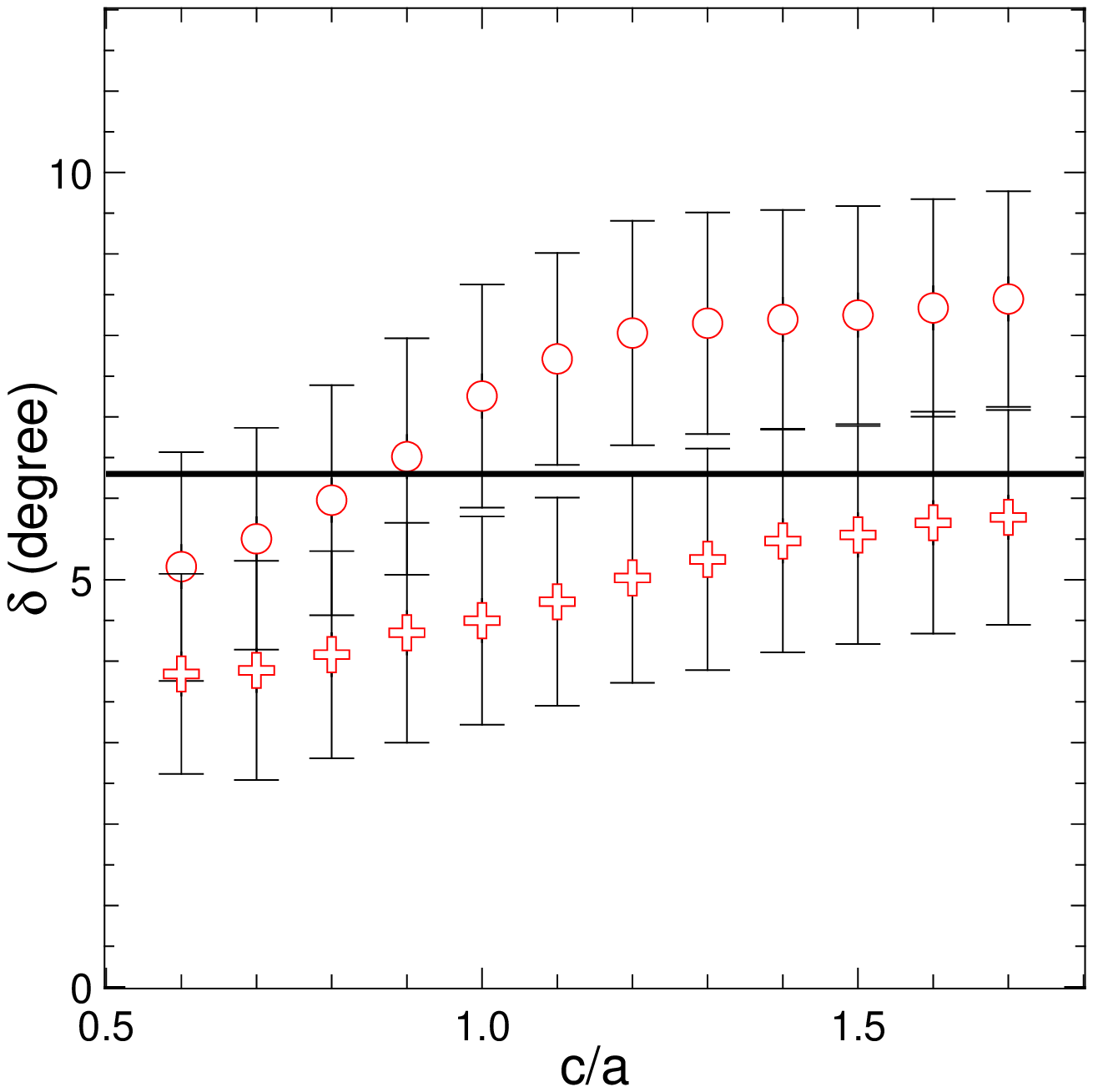}
\caption{Left panel: inclination of the  velocity ellipsoid as a function of
  the halo flattening $c/a_{\rho}$ in the RAVE selection function.  The full
  and thin dashed--dotted horizontal lines correspond to our measurement and
  error bars using  isotropic error laws.  The two  dashed curves correspond
  to the class of model I (top)  and II (bottom) in BT08 for which we varied
  the halo  flattening.  Right panel:  same as left  panel but for  a direct
  measurement. The  horizontal line  is the direct  measurement of  the tilt
  without  correction for the  velocity error  anisotropy.  The  circles and
  crosses are for Monte Carlo realisation  of the RAVE sample using the RAVE
  velocity error laws  for the class of model I  (circles) and II (crosses).
  The   error  bars  are   the  standard   deviation  obtained   from  1~000
  realisations.}
\label{f:ca}
\end{figure*}

The resulting measurements are presented in Fig.~\ref{f:ca} left panel where
the full  horizontal line is  our measurement from  Section~\ref{s:tilt} and
the  horizontal dashed  lines are  the 1$\sigma$  limit.  The  two remaining
curves correspond to the measurements obtained from our orbit analysis.  The
top curve is our prediction for the class of models I of BT08 and the bottom
curve for the  class of models II. For comparison,  the right panel presents
for the  direct measurement,  without correcting the  tilt for  the velocity
error  anisotropy.  The  circles  and  crosses  correspond  to  Monte  Carlo
realisations of  the RAVE  sample using the  orbit libraries where  the RAVE
velocity errors have been applied on the orbit library directly. Circles and
crosses are respectively  for the class of models I and  II.  The error bars
are the standard deviation of 1~000 realisations.\\

The classes of models show  the same general gross properties. The predicted
tilt rises  as the flattening decreases,  reaching a maximum  in the prolate
halo region. This maximum is expected and varies depending on the details of
each model.  It is due to  the fact that, when $c/a_\rho$ becomes large, the
potential becomes  separable in cylindrical coordinates. On  the other hand,
if $c/a_\rho$ approaches 0, the  problem reduces to the plane--parallel case
and the expected inclination at 1~kpc is $\delta\simeq3^{\circ}$.\\

The two  classes of models provide  different estimates for the  tilt of the
velocity  ellipsoid within  the limits  of our  sample.  The  tilt variation
versus $c/a_\rho$ is larger for prolate models than it is for oblate models.
For a density  flattening of 0.6, the difference  is only 2$^{\circ}$, while
for slightly prolate models the  difference reaches up to 5$^{\circ}$. Also,
in the oblate case, the expected inclination rises more quickly than for the
prolate case.  If one compares the predictions for the two classes of models
to the measured  inclination, a clear tendency is  present: the more massive
the halo is, the  more prolate it must be to match  the tilt of the velocity
ellipsoid   at   1~kpc.   The   two   extreme   cases  ``separate''   around
$c/a_\rho\simeq0.9$: while for a massive disc $c/a_\rho\leq0.9$ is necessary
to reproduce the  tilt, $c/a_\rho \geq 0.9$ is needed for  a massive halo in
the 1--$\sigma$  limit.  More specifically, the measured  orientation of the
velocity ellipsoid  is consistent with a  short scale length of  the disc if
the halo is oblate, while in the other case -- a scale length of the disc of
the order of 3~kpc -- the measured  value for the tilt implies that the halo
must be prolate.\\

Using  a direct  measurement, applying  the velocity  errors on  Monte Carlo
realisation of the  RAVE sample has the benefit of reducing  the errors by a
factor  $\sqrt 2$  (Fig.~\ref{f:ca}  right panel).   Nevertheless, the  bias
increases with the inclination,  see Fig.~\ref{f:test}, which results in the
difference  between the  two  models being  lower.  This direct  measurement
indicates that  low values of  $c/a_{\rho}$ are marginaly  inconsistent with
the measured  tilt, with  $c/a_{\rho}>0.7$ being prefered  even so  the same
general conclusions hold for both  modeling technique.  We note however that
applying the errors on the  orbit library (direct method) and correcting the
anisotropy of  the velocity errors (unbiased  measurement) produces slightly
different predictions for the tilt. This  is partly due to the fact that the
$U$  and $V$ velocities  are computed  from the  knowledge of  distances and
proper motions.  Hence, $U$ and  $V$ errors increases with distance which is
not  taken  into  account in  the  Monte  Carlo  simulation for  the  direct
measurement, the RAVE sample being  too small to estimate properly the error
laws as a function of distances. This effect is also reduced in the unbiased
measurement but is  still present since the correcting term  is added to the
true  error, hence distant  stars will  still have  on average  large errors
while nearer objects will have on average smaller errors.

Further constraints  on the minor--to--major  axis ratio are  also available
from independent studies.  For example,  the flattening of the dark halo has
been estimated from the shape of the Sagittarius dwarf tidal stream. A value
$c/a_\rho>0.7$,  with  a   preferred  flattening  of  $c/a_\rho\simeq1$,  is
obtained by  \citet{ib01} and \citet{maj03} using  respectively carbon stars
and  M--giants  from  the  2MASS  survey along  the  orbit  of  Sagittarius.
\citet{jo05}  gives  even   stronger  constraint  $0.75<c/a_\rho<1.1$  at  a
\mbox{3--$\sigma$} level, with oblate haloes strongly favoured if precession
of Srg's orbit is considered. In contrast, \citet{helmi04} and \citet{ljm05}
demonstrated  that  only  Galactic   potentials  with  prolate  halos  could
reproduce the  velocity trends in the  leading debris with  a preferred axis
ratio  $c/a_\rho=5/3$.  However,  \citet{ljm05} explored  a wide  variety of
Galactic potentials but failed to find  a single orbit that can fit both the
velocity trends and the sense of precession.

Looking back  at the solar neighbourhood,  if the Sgr stream  is orbiting in
oblate and  spherical potentials, Law  et al.  (2005) and  \citet{md07} both
predict that  the Sun is currently bathing  in a stream of  debris from Sgr,
passing  both inside  and  outside  the solar  circle.   Models orbiting  in
prolate potentials are on the  other hand inconsistent with this prediction.
\citet{be06},  \citet{newberg06} and  \citet{seabroke08} all  provide strong
evidence  for  the  absence  of  Sgr  debris  in  the  solar  neighbourhood.
\citet{fel06} argue that the origin of  the bifurcation in the Sgr stream is
only possible if  the halo is close to spherical,  as the angular difference
between the  branches is a measure  of the precession of  the orbital plane.
This suggests that the absence of  the Sgr stream near the Sun is consistent
with  nearly  spherical  and   prolate  Galactic  potentials  and  seemingly
inconsistent  with  oblate   potentials.   However,  recently  \citet{rpt07}
studied the Magellanic  System -- Milky Way interaction  using test particle
simulations  and compared  them  to HI  observations.   They concluded  that
$c/a_\rho<1$ values (oblate  halo) are prefered and allow  a better match to
HI observations.\\

In    Fig.~\ref{f:ca},    the   preferred    region    by   most    studies,
\mbox{$0.75<c/a_\rho<1$},  does  not permit  us  to  set strong  constraints
either on the flattening nor on the mass of the disc.  The measured value of
the tilt falls between the two  classes of models in the allowed region, and
the  error bars on  the RAVE  measurement do  not permit  us to  tighten the
parameter  space  reliably. For  strongly  prolate  halos,  as suggested  by
\citet{helmi04}, the class of models  II is preferred while short disc scale
length are marginally  rejected at the 2-$\sigma$ level.   If one adopts the
axis ratio  $c/a_\rho=1$ as preferred by \citet{maj03}  or \citet{ib01}, the
value of the tilt is better recovered with a model whose scale length of the
disc lies in the range $R_d=[2.5-2.7]$~kpc. Nevertheless, at the 1--$\sigma$
level, large and short values for $R_d$ are permitted with this analysis.\\

Various studies  in the  literature have used  star counts to  constrain the
scale length  of the thin  disc.  For example  recently Juric et  al. (2008)
measured  the scale  length  of  the stellar  disc  and found  $R_d=2.6$~kpc
($\pm20\%$) using  SDSS data.  Similarly,  using data from the  Bologna open
cluster  survey (BOCCE),  \citet{ci08} found  a  scale length  in the  range
2.25-3~kpc and  \citet{oj01} found $2.8$~kpc  using the 2MASS  survey. These
values are in good agreement with our finding but have similarly large error
bars. Refining our measurement will provide an independent constraint on the
scale length of the stellar disc.\\

%%%%%%%%%%%%%%%%%%%%%%%%%%%%%%%%%%%%%%%%%%%%%%%%%%%%%%%%%%%%%%%%%%%%%%%%%%%
%
%
%%%%%%%%%%%%%%%%%%%%%%%%%%%%%%%%%%%%%%%%%%%%%%%%%%%%%%%%%%%%%%%%%%%%%%%%%%%
\section{Conclusions} 

We measured  the tilt of the  velocity ellipsoid at  $\simeq$1~kpc below the
Galactic  plane  using a  sample  of  red clump  giants  from  the RAVE  DR2
catalogue. We find its inclination to be 7.3$\pm$1.8$^{\circ}$. Estimates of
the effect of contamination by  foreground stars and substructures have been
shown to be small and their effect on our measured value can be neglected.

We compared this value to predictions  from two extreme cases of mass models
for the  Milky Way proposed by  BT08. In the case  of a massive  disc with a
small  scale length  ($R_d=2$~kpc), the  inclination is  compatible  with an
oblate halo whose  minor--to--major axis ratio $c/a_\rho$ is  lower than 0.9
at the \mbox{1--$\sigma$} level. On the other hand, in the case of a massive
halo  with large disc  scale length  ($R_d\simeq3$~kpc), prolate  haloes are
prefered  with $c/a_\rho \geq  1$. When  a direct  measurement is  used, low
values  for  $c/a_{\rho}$  can   be  marginally  rejected,  indicating  that
$c/a_{\rho}>0.7$.

When further  independent constraints from previous  studies are considered,
we   find  that   an  intermediate   value   for  the   disc  scale   length
$R_d\simeq[2.5-2.7]$~kpc is  preferred for a  nearly spherical halo,  but no
extreme model  can be clearly ruled out,  due to our large  error bars. This
range is in good agreement with other studies relying on star count analysis
and deep  photometric surveys. Nevertheless  these results have  large error
bars of  the same  order as our  measurement and  cannot be used  to further
constrain the mass distribution. \\

RAVE continues  to acquire spectra and  this work relies on  the second data
release of the survey. So far  RAVE has collected more than 200~000 spectra,
4 times the  size of the sample used here. With  the current observing rate,
we can expect to multiply by ten  the size of our sample in the coming years
which will allow  us to significantly reduce our error bars.   By the end of
the survey, we  will be able to provide  a new mass model for  the Milky Way
galaxy with a constrained scale length of the disc and minor--to--major axis
ratio of the dark halo.

%%%%%%%%%%%%%%%%%%%%%%%%%%%%%%%%%%%%%%%%%%%%%%%%%%%%%%%%%%%%%%%%%%%%%%%%%%%
\section*{Acknowledgements}
AS would like to thank the anonymous referee for a very constructive report
and his comments that helped clarify this paper.

Funding for RAVE  has been provided by the  Anglo-Australian Observatory, by
the Astrophysical Institute Potsdam,  by the Australian Research Council, by
the German  Research foundation, by the National  Institute for Astrophysics
at  Padova, by  The Johns  Hopkins University,  by the  Netherlands Research
School  for Astronomy,  by  the Natural  Sciences  and Engineering  Research
Council of Canada,  by the Slovenian Research Agency,  by the Swiss National
Science  Foundation,   by  the  National  Science  Foundation   of  the  USA
(AST-0508996), by  the Netherlands Organisation for  Scientific Research, by
the Particle Physics  and Astronomy Research Council of  the UK, by Opticon,
by Strasbourg Observatory, and by  the Universities of Basel, Cambridge, and
Groningen. The RAVE web site is at www.rave-survey.org.

%%%%%%%

%%%%%%%%%%%%%%%	

%_______________________________________________________
%
% BIBLIOGRAPHIE
%_______________________________________________________
\bibliographystyle{aa}

%\appendix

\end{document}